\documentclass[useAMS,usenatbib]{mn2e}

\usepackage{times}
\usepackage{amsmath}
\usepackage{graphicx}  % Add graphics capabilities
\usepackage{epstopdf} % to include .eps graphics files with pdfLaTeX
\usepackage{url}

\newcommand{\myemail}{roskar@physik.uzh.ch}

\title{The Effects of Radial Migration on the Vertical Structure of
  Galactic Discs}

\author[R. Ro\v{s}kar et al.]{Rok Ro\v{s}kar$^{1}$\thanks{\myemail}, 
Victor P. Debattista$^{2}$,
Sarah R. Loebman$^{3}$
\\
$^1$Institute for Theoretical Physics, University of Z\"{u}rich, 
Winterthurerstrasse 190,
CH-8057 Z\"{u}rich, Switzerland\\
$^2$Jeremiah Horrocks Institute, University of Central Lancashire, 
Preston, PR1 2HE, UK\\
$^3$Astronomy Department, University of 
Washington, Box 351580, Seattle, WA 98195, USA\\
}

\begin{document}

\maketitle

\begin{abstract}

  We present evidence that isolated growing discs, subject to internal
  spiral perturbations, thicken due to both heating \emph{and} radial
  migration. We show this by demonstrating that the thickness and
  vertical velocity dispersions of coeval stars depend on their age as
  well as the change in their radii. While the disc thickens due to
  internal processes, we find that this induces only a minor amount of
  flaring. We further demonstrate the consequences of such thickening
  on the structural properties of stellar populations and find that
  they qualitatively agree with recent studies of the Milky Way disc.

\end{abstract}

\section{Introduction}
\label{sec:introduction}

Stars born at one radius within a disc galaxy such as the Milky Way
(MW) need not remain at that radius indefinitely, even if they retain
nearly circular orbits.  \citet{Sellwood:2002} have shown that the
interaction of stars with transient spirals at their corotation
resonance can lead to substantial changes in angular momentum of the
stars without significant heating.  This radial migration has many
important consequences for properties of stellar populations, such as
giving rise to a flattened age-metallicity relation with substantial
scatter \citep{Sellwood:2002, Roskar:2008a, Schonrich:2009}.

While it is typically assumed that the vertical and in-plane motions
of stars are largely decoupled, it is becoming increasingly apparent
that some coupling exists and that it affects the detailed
chemodynamical structure of discs. For example, as stars migrate
outwards they experience a smaller vertical restoring force and are
therefore expected to reach greater height above the mid-plane.
\citet{Schonrich:2009,Schonrich:2009a} explored this idea by using
analytical chemical evolution models that allowed stars to radially
migrate and assumed that in the process of migration their 
  velocity dispersions were conserved. They showed that under these
assumptions it was possible to create a thick disc entirely from
migrated stars that shared many of the chemodynamical properties of
the thick disc in the MW.

Disc thickening due to migrating stars has also been studied in
simulations by \citet{Loebman:2011}, who showed that outwards
migrating stars end up in the more vertically extended component of a
double $\mathrm{sech}^2$ profile density fit.  Similar to
\citet{Schonrich:2009a}, they found that subdividing the populations
by either kinematics or chemical abundances gives rise to an
artificial thin/thick disc dichotomy. Thus, radial migration appears
as yet another possible mechanism for creating a thickened disc
component, but because it relies entirely on internal processes it is
conceptually orthogonal to most other thick disc formation scenarios.

Other mechanisms for creating a thickened disc component involve
either heating the disc through bombardment
(e.g. \citealt{Quinn:1993}), forming a thickened disc from the
remnants of a gas-rich last major merger \citep{Brook:2004} or
accreting an extra-planar stellar component \citep{Abadi:2003}. All
three of these mechanisms are a result of the cosmological environment
and consequently the thick disc is often considered to be an indicator
of the MW's cosmic history. On the other hand, a thickened component
built up by migration has little to do with cosmological evolution but
depends largely on internal disc dynamics. If migration can influence
the properties of the thicker stellar component then it can muddle the
signatures of cosmologically-relevant events that undoubtedly marked
the early stages of the MW's disc formation. The formation of the
thick disc through heating by clump instabilities is in principle also
an in-situ formation mechanism \citep{Bournaud:2009}, but it still
relies on the cosmological environment for the rapid gas accretion.

\citet{Sales:2009a} proposed using the eccentricity distribution of
thick disc stars to distinguish between thick disc formation
mechanisms.  Consequently, several groups sought to apply the
\citet{Sales:2009a} eccentricity test to observational
data. \citet{Dierickx:2010}, \citet{Casetti-Dinescu:2011}, and
\citet{Wilson:2011}, all reached a general conclusion that the
eccentricity distribution is mostly inconsistent with the accretion
scenario, but broadly consistent with the other three (the gas-rich
merger scenario being the most favorable). The eccentricity
distribution is not alone enough to break the degeneracies among the
models, but such constraints become stronger once metallicities and
abundances are also considered: \citet{Liu:2012} showed that low
eccentricity stars from a SEGUE survey \citep{Yanny:2009} sample
follow a continuous distribution in the [Fe/H]-[$\alpha$/Fe] plane
while the high eccentricity stars seem disconnected in this plane.  

The separation in the [$\alpha$/Fe] vs. [Fe/H] plane of thick- and
thin-disc selected stars lends strong support to the idea that the two
discs are formed as separate structures \citep{Bensby:2003}. In such a
formation scenario, the variation of structural parameters in
[alpha/Fe] vs. [Fe/H] for two distinct populations would be expected
to be abrupt even for an unbiased sample of stars. However, recently
\citet{Bovy:2012b} found that such a strong dichotomy does not exist
for a selection function-corrected sample of SEGUE
stars. \citet{Bovy:2012b} find a smooth distribution of structural
trends when stars are separated into mono-abundance populations with
alpha-rich, metal-poor large scale height populations having a short
scale-length, while metal-rich stars have low scale-heights and long
scale-lengths. (a similar result was found for the high-$\alpha$ SEGUE
population by \citealt{Cheng:2012}). These findings are consistent
with \citet{Bensby:2011} who also found a short scale-length for
targeted thick-disc stars using entirely different data. These recent
results contrast strongly with the usual view that the thick disc has
a larger scale height and a longer scale length, as is obtained, for
example, by fitting the stellar density with a two-component model
\citep{Juric:2008}.

The stars in individual abundance bins are also vertically isothermal
\citep{Bovy:2012c}, strengthening the argument that they are single
populations.  Both of these properties seem to favour strong internal
evolution for the MW disc, because if a disc builds up entirely from
internal processes, a distinct second component that dominates away
from the plane does not readily form. It is, however, also possible to
recover many of these trends in the cosmological context if the disc
initially forms hot and thick, subsequently forming increasingly
thinner populations as its self gravity gradually increases
\citep{Bird:2013,Stinson:2013}.

However, even if stars form in a thicker component early in the
history of the disc, they do not become immune to disc perturbations.
$N$-body simulations have shown that stars in the thick disc were also
able to migrate though to a lesser extent \citep{Solway:2012}. The
same study showed that migrating stars on average (but not
individually) conserve their vertical action, rather than their
vertical energy.

Recently, \citet{Schonrich:2012} studied the co-dependence of vertical
and horizontal motions analytically. They found that the vertical
motion significantly affects the detailed in-plane velocity
distributions. Following \citet{Binney:2010}, the coupling between the
two components of motion was achieved through the assumption of
vertical action invariance.  This assumption has been shown to be a
reasonable approximation by more rigorous torus modeling
\citep{Binney:2011b}. While the above analytic studies have focused on
the co-dependence of oscillations about the guiding centre and the
vertical motion, in this Paper we explore the implications of radial
migration (i.e. the change in the guiding centres) for the vertical
distribution. Under action conservation in a radially-increasing disc
potential, any change in radius will lead to a modification of the
vertical motion. However, while stars oscillate in radius by up to
$\sim 1-2$ kpc they might migrate radially by many kpc and we can
therefore expect that migration would have a larger effect on their
vertical motion.

However, \citet{Minchev:2012a} argue that vertical action conservation
\emph{prevents} the stars from thickening as they migrate
outwards. They support this claim with results from isolated
$N$-body+SPH simulations from the GalMer database
\citep{Chilingarian:2010} as well as sticky-particle semi-cosmological
simulations \citep{Martig:2012a}. Their conclusion disagrees sharply
with the findings of \citet{Schonrich:2009a}, who assumed
\emph{energy} conservation rather than action conservation when
modeling the vertical distribution in their analytic models. However,
it also disagrees with \citet{Loebman:2011} who used high-resolution
$N$-body/SPH simulations and found that the migrated population formed
a thicker component that was shown to be broadly consistent with MW
thick-disc trends in both chemistry and kinematics.

In \citet{Roskar:2012}, we focused on the details of the radial
migration mechanism in a suite of $N$-body/SPH simulations. We found
evidence that the largest migrations were taking place at the
corotation resonance (CR) of dominant spirals, confirming the
mechanism proposed by \citet{Sellwood:2002}. In this paper, we
elucidate the connection between radial migration and vertical disc
structure using the fiducial simulation of \citet{Roskar:2012}. To
gain insight into the physical nature of the processes involved, we
compare the simulation data with expectations based on the simplified
analytic models of \citet{Schonrich:2012}. The paper is organized as
follows: our methods are presented in Section~\ref{sec:sims}; the
basic results are presented in Section~\ref{sec:results}; we compare
the results to a simple analytic model for disc thickening in
Section~\ref{sec:model}; we discuss the consequences of such
thickening on the trends in structural properties of stellar
populations in Section~\ref{sec:trends}.

\section{Simulations}
\label{sec:sims}
We analyze the same $N$-body/SPH simulation that we used in
\citet{Loebman:2011}. This simulation is a re-run of the fiducial
simulation analyzed in \citet{Roskar:2008, Roskar:2008a,Roskar:2012}
using metal diffusion, but is otherwise identical in every
respect. The initial conditions consist of two NFW
\citep{Navarro:1997} halos, one of dark matter with a mass of
$10^{12}~M_{\odot}$ and the other of gas in hydrostatic equilibrium
with a mass of $10^{11}~M_\odot$, both sampled with $10^6$
particles. The gas also rotates with $\lambda = 0.039$ and $j\propto
R$, where $j$ is the specific angular momentum and $R$ is the
cylindrical radius. We use the code GASOLINE \citep{Wadsley:2004} to
perform the computation. Once the simulation begins, the gas cools and
collapses into a disc that forms stars according to a standard
prescription \citep{Stinson:2006} with a density threshold of
0.1~amu/cc and a temperature cutoff of $1.5 \times 10^4$~K. Star
particles provide ``feedback'' to the gas by injecting it with energy
and polluting it with supernova ejecta. The simulations are evolved
completely in isolation and no cosmological context is included. By
the end of the simulation, the disc is populated by more than
$2\times10^6$ star particles. The softening we use for the baryonic
component is 50~pc. The reader is referred to \citet{Roskar:2008} and
\citet{Roskar:2012} for further details on the simulations. The
simulations were analyzed with the aid of the \textsc{Python}-based
analysis package
\textsc{Pynbody} \citep{pynbody}\footnote{\url{http://code.google.com/p/pynbody/}} and IPython \citep{Perez:2007}.

Our simulation code includes prescriptions for the generation of
metals in supernova type Ia and II explosions as well as in AGB
stars. SN II yields are taken from \citep{Raiteri:1996}, SN Ia yields
from \citet{Thielmann:1986}, and the mass returned to the ISM via
stellar winds follows \citet{Weidemann:1987}. We can therefore follow
the abundances of $\alpha$ elements relative to Fe (Oxygen is used as
the $\alpha$ element proxy). Due to uncertainties in metal yields, the
absolute values of [O/Fe] do not match with observations, but relative
trends are nevertheless informative. For details on the metal
enrichment implementation see \citet{Stinson:2006}. As in
\citet{Loebman:2011}, the simulation we use here also implements a
prescription for the diffusion of metals through the ISM.

\section{Results}
\label{sec:results}

\begin{figure}
\centering
\includegraphics[width=3.in]{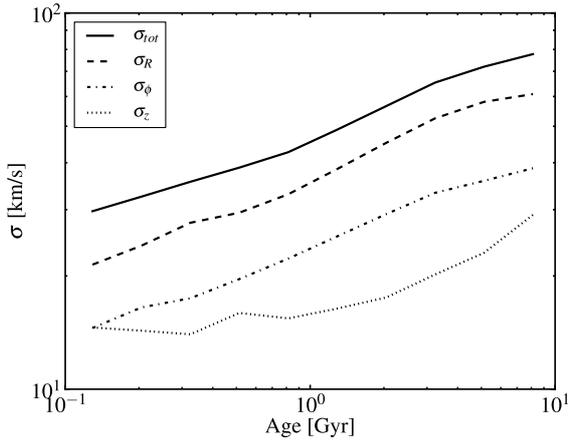}
\caption{Age-velocity dispersion plot for stars with $7.5 < R
  \mathrm{~[kpc]} < 8.5$ and $|z| < 200 \mathrm{~[pc]}$ after 10 Gyr
  of evolution.}
\label{fig:agesigma}
\end{figure}

The simulation yields a disc that is kinematically similar to that of
the Milky Way. In Fig.~\ref{fig:agesigma} we show the age-velocity
dispersion relation for stars that are found in the ``solar
neighborhood'' at the end of the simulation. We define solar
neighborhood as a disc region between 7.5-8.5 kpc and within 200 pc
from the plane. The disc scale length in our model is $\sim2.5$~kpc,
so this is a reasonable approximation to the region of the MW disc
where analogous relations have been observed. The age-velocity
relation from the Geneva-Copenhagen survey \citep{Holmberg:2009} shows
similar power-law dependencies and values, though in our case the
oldest stars are hotter by $\sim 10\%$. The vertical velocity
dispersion shows a shallow dependence for the young stars, similar to
that shown for the GCS sample \citep{Seabroke:2007}. Part of the
increased dispersion for the old stars may be due to the fact that our
simulation yields a slightly more massive disc (the circular velocity
$v_c \sim 250$~km/s). Fitting power-laws to the relations in
Fig.~\ref{fig:agesigma} we obtain power-law indices of 0.24, 0.26,
0.25, and 0.15 for $\sigma_{tot}$, $\sigma_R$, $\sigma_{\phi}$, and
$\sigma_z$ respectively. These indices are lower than those observed
in the MW, indicating that an additional source of heating may be
needed. On the other hand, the fact that the heating indices are not
too high also means that the amplitudes of non-axisymmetric structure
forming in the disc are not unreasonably high. A related issue is that
the youngest stars are not quite as kinematically cool as the
observations due to the dispersion floor in the gas component, which
also contributes to higher overall velocity dispersions. We discuss
this last point further in Sect.~\ref{sec:model}.

\subsection{Thickening and vertical velocity dispersion due to migration and heating}

Because radially migrating stars feel a smaller restoring force
towards the midplane as they move outward, the amplitude of their
vertical oscillations increases. However, at the same time, heating
processes due to the recurring spiral structure also heat the stars,
regardless of whether they have migrated or not, providing another
source of increased vertical motion. Comparison between Figs.~3 and~5
in \citet{Loebman:2011} shows this implicitly.  Their Fig.~3 shows
that when significant migration is present, the stars dominating the
distribution away from the plane have migrated from the interior of
the disc. By contrast, in the model with less migration shown in their
Fig. 5, the populations away from the plane are there mostly because
of heating (i.e. they are old) rather than migration (they formed
in-situ).

\begin{figure*}
\centering
\includegraphics[width=7in]{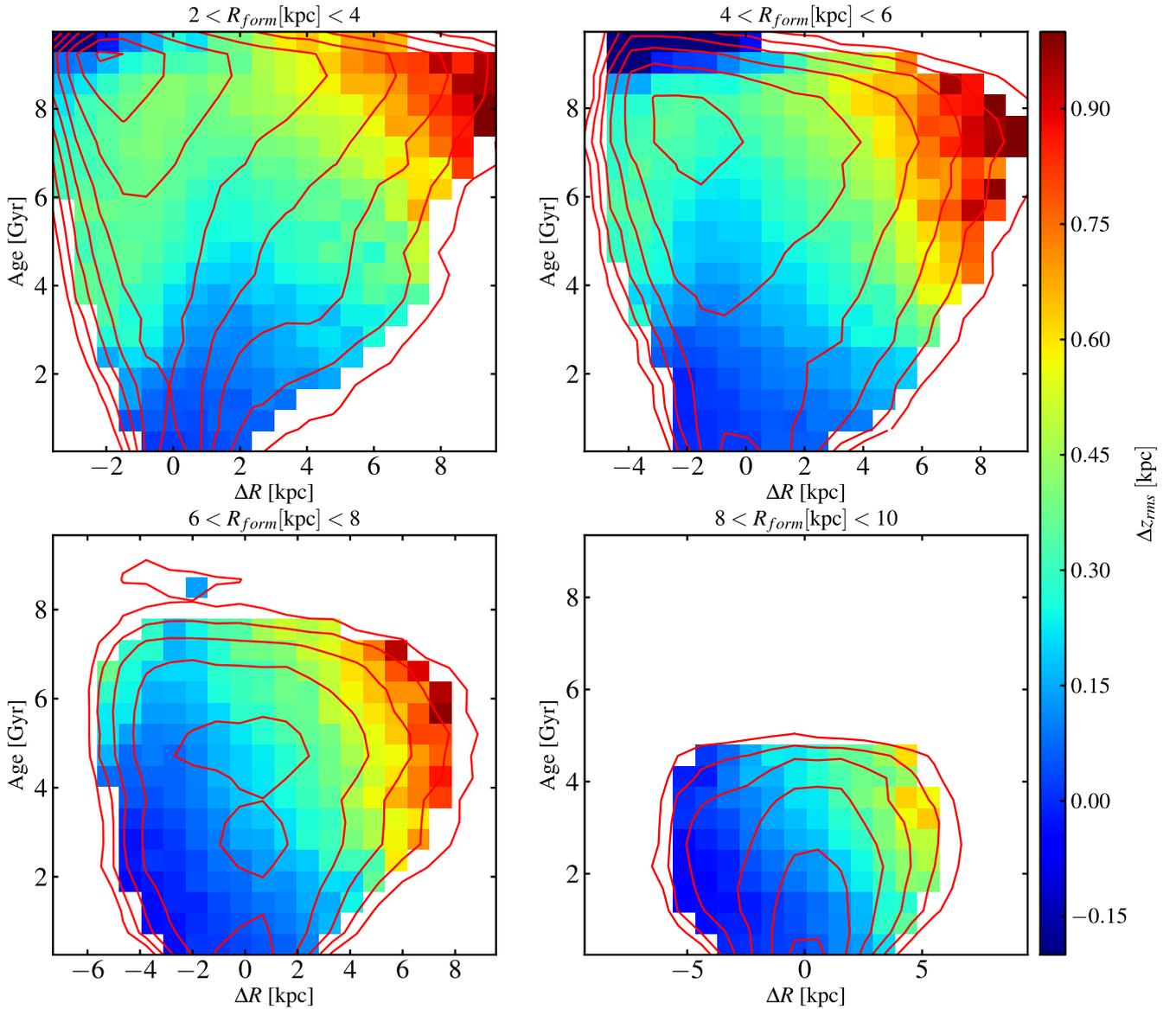}
\caption{Mean change in thickness as a function of age and $\Delta R$
  plotted in four different bins of formation radii. Contours show the
  particle mass density and are logarithmically spaced from 10 to
  $10^4$ particles per bin. Colours correspond to the change in
  thickness, defined as the root-mean-square of the vertical
  displacement from the midplane, $z_{rms} =
  \sqrt{\bar{z}^2}$. Increase in thickness is a function of both, age
  and $\Delta R$, i.e. stars undergo heating but their vertical
  distribution is also affected by the effective change in potential
  as they migrate inwards or outwards in the disc. Note that the
  $x$-axis range is different in each panel.}
\label{fig:hz}
\end{figure*}

In Fig.~\ref{fig:hz} we show the relative dependence of stellar
population thickness on radial migration and heating. As a measure of
thickness, we use the model-independent $z_{rms} \equiv
\sqrt{\bar{z}^2}$. Each panel shows the distribution of $\Delta
z_{rms} = z_{rms,now} - z_{rms,form}$, i.e. the change in thickness
since birth, as a function of age and $\Delta R \equiv R_{now} -
R_{form}$, for a given range of formation radii. The distributions
shown in each of the panels are made by selecting particles in a
particular range of \emph{formation} radii (indicated at the top of
each panel).  We bin the particles in this space on a grid -- each
bin then corresponds to roughly a coeval population (born at
approximately the same radius, of the same age, and migrated by the
same amount). The contours show the mass density of particles.

If all the changes to the thickness of each population were due to
heating, we would expect age to be the determining factor in setting
the $z_{rms}$. As a result, the 2D gradient in $\Delta z_{rms}$ would
be predominantly in the vertical direction. Conversely, if all the
changes in thickness were due to radial migration, the gradient would
be horizontal.

\begin{figure*}
\centering
\includegraphics[width=7in]{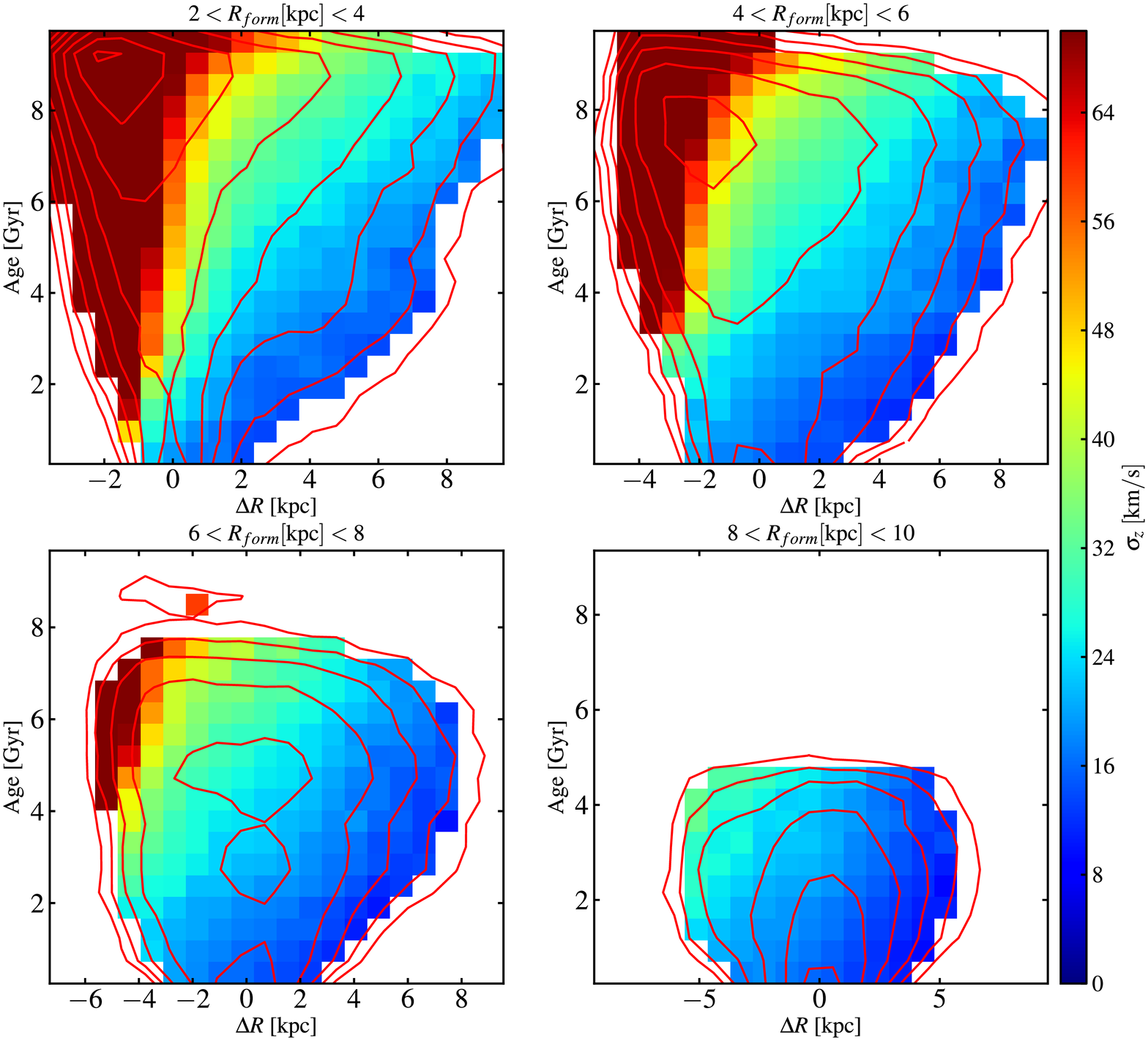}
\caption{Same as Fig.~\ref{fig:hz} except here the colours correspond
  to $\sigma_z$, the vertical velocity dispersion at the end of the
  simulation. The old stars that the ``solar neighborhood''
  (i.e. $\sim8$~kpc) from the interior of the disc have $\sigma_z \sim
  40-50$~km/s in broad agreement with thick disc values in the MW.}
\label{fig:sigmaz}
\end{figure*}

The gradients in all of the panels of Fig.~\ref{fig:hz} are
\emph{diagonal}, with $\Delta z_{rms}$ increasing in the direction of
older ages and larger $\Delta R$, implying that radial migration as
well as heating contribute substantially to the vertical thickness of
the stellar populations. This is true for all ages and all $\Delta
R$. For positive $\Delta R$, $\Delta z_{rms}$ is a monotonic function
of $\Delta R$.  Stars migrating inward decrease their $z_{rms}$ so
long as they stay away from the central bulge region. Those stars that
migrate to the very centre of the disc increase their $z_{rms}$,
presumably due to rapid heating that occurs there due to the presence
of multiple inner Lindblad resonances and a weak oval that develops in
the centre from time to time. The high velocity dispersions for these
stars indicate that this may indeed be the case (see
Fig.~\ref{fig:sigmaz} discussed below). The oldest stars in the
interior of the disc (upper-left corners of the top two panels) are on
eccentric orbits and therefore not as affected by disc perturbations
but their $z_{rms}$ decreases due to adiabatic contraction. Our main
finding from Fig.~\ref{fig:hz} is that 1) at any given age, thickness
is a monotonic function of $\Delta R$, and 2) at any $\Delta R$
thickness is a monotonic function of age. 

We find a very similar kind of co-dependence for $\sigma_z$, the
vertical velocity dispersion, shown in Fig.~\ref{fig:sigmaz}. The
velocity dispersions of the stars migrating outward are \emph{lower}
than the dispersions of those migrating inwards. This is to be
expected because the stars are moving to a region of a shalower
midplane potential and because stars in the inner disc heat very
efficiently (due to inner Lindblad resonances or non-axisymmetric
structure). Nevertheless, the dispersion clearly depends on both
parameters, age as well as distance migrated since birth, consistent
with Fig.~\ref{fig:hz}.

In Figs.~\ref{fig:hz} and~\ref{fig:sigmaz} we have shown that the
thickness and vertical velocity dispersion of coeval stellar
populations depend on the magnitude of migration as well as their
age. In other words, at a fixed present-day radius, the velocity
dispersion and thickness of a population depends on its birth location
\emph{and} its time of formation. However, we have also shown that
$\sigma_z$ of stars actually \emph{decreases} as they migrate outward,
so can these stars entering the solar neighborhood from the inner disc
then still masquerade as the thick disc? From Fig.~\ref{fig:sigmaz},
we can see that the old ($> 8$~Gyr) stars that have migrated outward
by several kpc have vertical velocity dispersions of 40-50 km/s. These
stars comprise the $\alpha$-old population that in the MW has very
similar velocity dispersions \citep{Liu:2012,Bovy:2012c}; note that
for the purposes of \emph{defining} the thick disc, the
``characteristic'' $\sigma_z$ is taken to be $\sim 35$~km/s
\citep{Bensby:2003}, a criterion met by this migrated population. Even
if the populations ``cool'' as they migrate outwards, their
present-day properties can remain consistent with a thickened,
$\alpha$-rich population. Note that in our models the only stars with
the appropriate $\sigma_z$ are $>5$~Gyr old and certainly the only
ones that are \emph{thickened} enough to represent the thick disc are
even older and have migrated substantially. The in-situ young stars
have $\sigma_z < 20$~km/s, making up the ``thin'' disc.

\citet{Minchev:2012a} similarly found that velocity dispersions
decrease as stars migrate outwards. They interpreted the decrease in
$\sigma_z$ as an indication that radial migration cannot contribute to
a thickened population. However, because the stars migrating from the
inner to outer disc begin with a relatively high velocity dispersion,
they still end up somewhat kinematically hotter relative to the
in-situ population when they arrive to the outer disc. Furthermore, as
Fig.~\ref{fig:hz} clearly shows, the outward-migrating populations
thicken vertically as well. We find therefore that it is indeed
possible to form a thickened and kinematically hotter component with
the aid of radial migration. Our results therefore appear
qualitatively similar to those of \citet{Minchev:2012a}, although we
arrive at different conclusions. We speculate further that possible
discrepancies between our findings and those of \citet{Minchev:2012a}
may arise in part due to the rapid heating apparent in their
simulations due to a violent initial instability. At later times, the
phenomenon driving the disc trends in their models (the same
simulations were also discussed in \citealt{Minchev:2011} and
\citealt{Minchev:2012}) is efficient angular momentum redistribution
during bar growth (e.g. \citealt{Hohl:1971}) as opposed to more
quiescent redistribution by spirals which dominates the evolution in
our simulations \citep{Roskar:2012}. Furthermore, we need to emphasize
here that while it is clear that the stars in our simulations (and all
others) heat vertically, the mechanism responsible for such heating in
$N$-body models is not well understood \citep{Sellwood:2013}.

\subsection{Disc flaring}
\label{sec:flaring}

An important consideration for any model that deals with the
thickening of the stellar disc is the change in scale height with
radius, i.e. disc flaring. While substructure bombardment is
particularly efficient at thickening discs, it may also cause rather
drastic flaring to occur \citep{Kazantzidis:2009}. Such flaring is not
typically seen in extragalactic observations of edge-on discs
\citep{degrijs:1996}.

\begin{figure}
\centering
\includegraphics[width=3in]{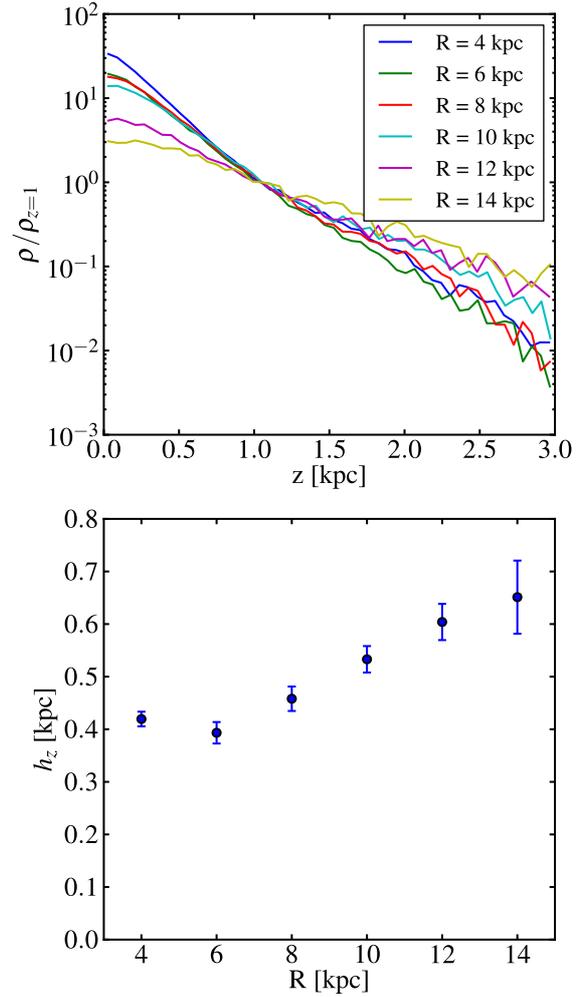}
\caption{{\bf Top:} Vertical density profiles at several different
  radii, normalized to their values at $z=1$. {\bf Bottom:} Scale
  heights of the thicker sech$^2$ component as a function of
  radius. From the inner disc to the outermost regions we find a
  modest increase in scale height.}
\label{fig:flaring}
\end{figure}

While, radial migration has also previously been associated with
dramatic flaring \citep{Minchev:2012a}, we show in
Fig.~\ref{fig:flaring} that this is not necessarily the case. In the
top panel, we show the vertical profiles of the disc at several
different radii, all normalized to their values at 1~kpc from the
plane to more easily compare the profiles furthest from the plane. The
vertical distribution clearly becomes more extended in the outer parts
of the disc, but the flaring is rather minor.

In the bottom panel, we quantify the flare by showing the values of
maximum-likelihood estimates (MLE) for the scale height of the thicker
component of a double sech$^2$ distribution as a function of
radius. We use MLE instead of a least-squares fit to the binned
vertical profiles because we found those fits to be unreliable, while
MLE gives consistently good results. To find the MLE parameters, we
maximize the log-likelihood function
\[
\ell(h_1,h_2,f|z) = 
\sum_{i=0}^N \mathrm{ln}\left[\rho(z_i|h_1,h_2,f)\right] - \mathrm{ln}(\rho_0),
\]
with 
\[
\rho(z)/\rho_0 = (1-f)\mathrm{~sech}^2(z/2h_1) + f\mathrm{~sech}^2(z/2h_2)
\]
where $\rho$ is the volume density, $N$ is the total number of
particles in the bin, $z_i$ is the vertical position of the $i$-th
particle, $h_1$ and $h_2$ are the thin and thick disc scale heights,
$f$ is the thick disc fraction, and $\rho_0$ is the midplane
density. To estimate the parameters, we select particles found in
1~kpc radial bins centred on each point shown in
Fig.~\ref{fig:flaring} and up to 4~kpc from the midplane. Following
\citet{Bovy:2012b}, the likelihood function $\ell$ is maximized using
an MCMC sampler \citep{Foreman-Mackey:2012}, yielding parameter
estimates and uncertainties, represented by the error bars in
Fig.~\ref{fig:flaring}.

The increase in scale height is only $50\%$ across approximately
10~kpc (corresponding to 4 disc scale lengths). The thick disc
fraction is $\sim 12-20\%$ in the main disc region (4-10~kpc) and
rises to $\sim40\%$ in the outer disc. Interior to 10~kpc, the scale
height is essentially constant. Note that most studies fitting the
vertical density of the Milky Way disc (e.g. \citealt{Gilmore:1983,
  Juric:2008, Bovy:2012b}) use an exponential fitting function and
that the scale heights between the two functional forms are not
strictly comparable. We find that fitting our simulated particle
distribution with an exponential model yields scale heights larger by
up to $\sim 30\%$, making them only slightly smaller than similar
thick disc scale height measurements in the MW. However, we use the
sech$^2$ because it gives a better fit to our model.

\section{Disc Thickening due to Conservation of Vertical Action}
\label{sec:model}

Here we consider the simplified 1D problem with a star oscillating
above and below the plane of a thin disc. This is similar to the model
described by \citet{Minchev:2012a}, who use it to show that the
vertical velocity dispersion is expected to decrease as stars migrate
outward if the vertical action is conserved. We show that this same
model leads to the conclusion that the stellar vertical distribution
\emph{must} thicken. In this simple case, the vertical frequency is
\begin{equation}
\label{eqn:nu}
\nu \propto \sqrt{\rho(R)},
\end{equation}
where $\rho(R) = \rho_0 e^{-R/R_d}$ is the midplane density of the
disc at radius $R$, $\rho_0$ is the density at the disc centre, and
$R_d$ is the disc scale length. The vertical action can be
approximated by $L_z = E_z/\nu$, where $E_z$ is the vertical
energy. For a population of stars we assume that $E_z \sim
\sigma_z^2$.  If we assume that the vertical action $L_z$ is conserved
\citep{Solway:2012} then 
\begin{equation}
\label{eqn:sigmaz}
\sigma_z^2 \propto \nu \propto e^{-R/2R_d},
\end{equation}
as shown previously by \citet{Minchev:2012a}.

The vertical density distribution of
an isothermal population is given by \citep{Spitzer:1942}
\begin{equation}
\label{eqn:rhoz}
\rho(z) = \rho_{_{z=0}} \mathrm{~sech}^2(z/2h_z),
\end{equation}
with the vertical scale-height 
\begin{equation}
\label{eqn:hz}
h_z = \frac{\sigma_z^2}{2 \pi G \Sigma(R)}.
\end{equation}
Plugging Eq.~\ref{eqn:sigmaz} into Eq.~\ref{eqn:hz}, and assuming that
the disc surface density $\Sigma \propto e^{-R/R_d}$, one finds that
$h_z \propto e^{R/2R_d}$, i.e. the scale height \emph{increases} with
positive changes in radius. Combining the approximation for change in
$\sigma_z$ for a migrating population with Eq.~\ref{eqn:hz} yields a
simple expression relating the change in radius $\Delta R$ and $R_d$
to the ratio of initial and final scale heights, $h_{z,f}/h_{z,i}$:
\begin{equation}
\frac{h_{z,f}}{h_{z,i}} = \mathrm{exp}\left(\frac{\Delta R}{2R_d}\right).
\label{eqn:dr_dz}
\end{equation}
This simple derivation of Eq.~\ref{eqn:dr_dz} ignores the presence of
other disc components; such an isothermal population is just one of
many that make up the disc. In particular, the addition of mass (by
gas accretion and star formation) in the midplane would adiabatically
compress the migrated population, which has previously been shown to
alleviate some of the thickening effects of infalling substructures
\citep{Villalobos:2010, Moster:2010b}. Furthermore, even in a disc not
subject to outside perturbations, $\sigma_z$ will also increase with
time due to heating from disc structure thereby further boosting the
scaleheight, an additional effect that we discuss below.

Eq.~\ref{eqn:dr_dz} is an approximation to the more general expression
derived by \citet{Schonrich:2012} (their Eq.[27]), which instead of
the factor 2 on the right hand side includes a factor of $(2 +
\alpha)$. The $\alpha$ changes depending on the magnitude of vertical
oscillation. Note that the change induced in the $z$ distribution due
to adiabatic invariance occurs irrespective of the reason for the
change in radius. The change may be due to radial oscillations about
the guiding radius \citep{Schonrich:2012} or due to the change in the
guiding radius itself. Therefore, the increase in $z_{rms}$ with
increasing $\Delta R$ in Fig.~\ref{fig:hz} is due to a complex
combination of radial heating, migration, and the evolution of the
galactic potential.

\begin{figure*}
\centering
\includegraphics[width=7in]{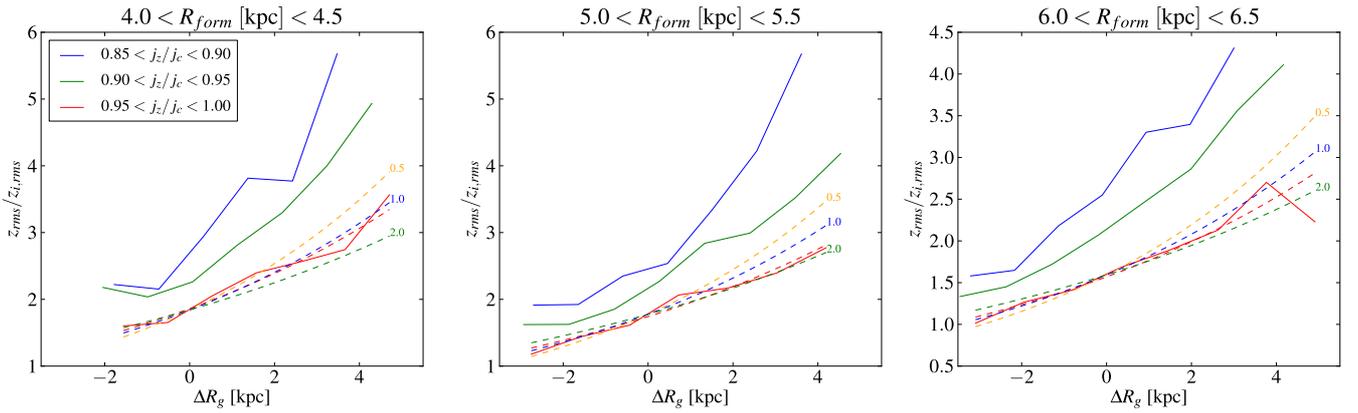}
\caption{The ratio of $z_{rms}$ (thickness now) to $z_{i,rms}$
  (thickness at formation) of 5-6 Gyr old stars that were formed
  within the indicated radial range, plotted against the change in
  guiding radius $\Delta R_g = \Delta j_z/V_c$ from their birth to the
  present, where $\Delta j_z$ is the change in angular momentum and
  $V_c$ is the circular velocity. The colours of the solid lines
  correspond to different ranges in the orbital circularity parameter
  $j_z/j_c(E)$. The overplotted red dashed lines are exponential fits,
  indicating that the changes in thickness satisfy the expectations
  from adiabatic invariance of the vertical action (see text). The
  other coloured dashed lines show the relation from Eq.[27] of
  \citet{Schonrich:2012} for the kinematically coldest population,
  overplotted for different values of $\alpha$ and assuming a scale
  length of 2.5 kpc.}
\label{fig:zrms_jzform}
\end{figure*}

In Fig.~\ref{fig:zrms_jzform} we try to separate out the dependence of
stellar population thickness on these processes. We focus on stars
that are 5-6 Gyr old and formed within the range of specified
radii. We then further separate these stars by their orbital
circularity parameter, $j_z/j_c(E)$ measured at the present time,
where $j_z$ is the specific angular momentum and $j_c(E)$ is the
maximum specific angular momentum for a star with a given energy. The
solid lines show the variation in fractional change in thickness
$z_{rms}/z_{i,rms}$ versus change in guiding radius since birth,
$\Delta R_g$ for different ranges in orbital circularity. The range in
$j_z/j_c$ is quite representative of this particular subset of stars,
i.e. there are not many particles within the chosen constraints that
are significantly kinematically hotter. We define $R_g$ in terms of
angular momentum, i.e. $R_g = j_z/V_c$, where $V_c = 250
\mathrm{~km/s}$ is the approximate circular velocity. The age and
$R_{form}$ ranges are chosen to give a reasonable sampling of the main
part of the disc.

Looking at Fig.~\ref{fig:zrms_jzform}, one can now identify the
dependence of a stellar population's vertical distribution on heating
and migration separately. For the kinematically coldest population,
the thickness changes by a factor of 3 along the range of $\Delta
R_g$. On the other hand, it increases by a further factor of $\sim2$
from the coldest population to the hottest population, indicating that
both processes actually contribute in similar amounts to the
thickening.

The evolution of the thickness of this coeval stellar population can
be described by the simple disc model discussed above under the
assumptions of adiabatic invariance. We extend the simple formula in
Eq.~\ref{eqn:dr_dz} to that derived by \citet{Schonrich:2012} and plot
the result in Figure~\ref{fig:zrms_jzform} for different values of
$\alpha$ shown with coloured dashed lines using a disc scale length
$R_d = 2.5$~kpc. The least-squares best fit is shown with the red
dashed line. Note that for a sech$^2$ distribution, $z_{rms}$ is
smaller than the scale height by approximately 10\%, so $z_{rms}$
gives a reasonable approximation to $z_0$ that we can use in
Eq.~\ref{eqn:dr_dz}.

The agreement between the simulation and the analytic prediction of
the simple model, given the assumptions, is encouraging. Note that the
likely value of $\alpha$ should be 1-2 since particles with the
largest values of $j_z/j_c$ have amplitudes of vertical oscillations
$| z_{max} | \ll R_d$ \citep{Schonrich:2012}, which provides an
excellent description of the simulation data. We have experimented
with using different criteria for the $R_{form}$ and age selection and
found no appreciable differences, though the best-fit parameters start
to deviate from the expected values close to the centre or far in the
outer disc where the surface density profile can no longer be
described as exponential. The three different ranges of $R_{form}$
shown in Figure~\ref{fig:zrms_jzform} show no appreciable
differences. If we choose a different age range, the relations stay
qualitatively the same.

\begin{figure*}
\centering
\includegraphics[width=7in]{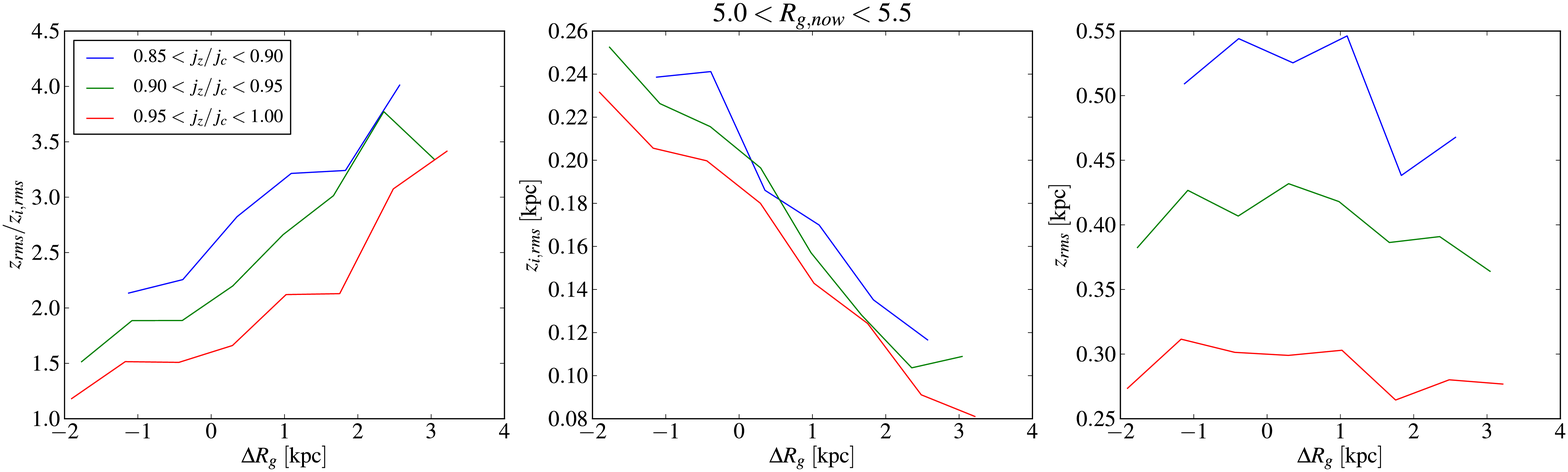}
\caption{The change in population thickness as a function of $\Delta
  R_g$, the change in guiding radius since birth, defined as $R_g =
  j_z/V_c$ (same as in Figure~\ref{fig:zrms_jzform}) for particles
  with $R_g$ in the indicated range. The left panel shows the ratio of
  $z_{rms}/z_{i,rms}$, predicted by Eq.~\ref{eqn:dr_dz} to follow an
  exponential. The centre and right panels show $z_{rms}$ at time of
  formation and at the present time respectively. Stars used in this
  figure are 5-6 Gyr old. }
\label{fig:zrms_jzfinal}
\end{figure*}

Heating by spirals in our simulation results in increased random
motion in-plane and out-of-plane. The increased thickness of the
populations on slightly more eccentric orbits shown with the blue and
green lines is therefore due to the combined effects of higher
vertical random energy as well as amplified vertical oscillations due
to adiabatic invariance at the orbital apocentres.

We have shown above that the thickness of a coeval population is an
exponential function of its migration in the disc. It is interesting
to pose a similar question for stars that share the same angular
momentum \emph{today} but that may have originated in different parts
of the disc, i.e. are not to be considered a coeval population. As in
Fig.~\ref{fig:zrms_jzform}, in the leftmost panel of
Fig.~\ref{fig:zrms_jzfinal} we show the ratio $z_{rms}/z_{i,rms}$ for
stars with $5 < R_{final} \mathrm{~[kpc]} < 5.5$ vs. $\Delta R_g$. We
see similar exponential trends at all values of $j_z/j_c$, as
before. Since the most circular (red line) population is the least
affected by heating, we can once again use it as a proxy for the
overall effect of radial migration on population thickness. Across the
range of $\Delta R_g$, we find that migration can lead to an thickness
increase by a factor of 2.5 (we cannot eliminate all heating so
thickness increases by 50\% simply by virtue of these stars spending 5
Gyr in the disc). Again, taking the range from blue to red as the
effect of heating, we see that heating contributes another factor of
$< 1.5$. Based on this and Fig.~\ref{fig:zrms_jzform}, \emph{we
  conclude that migration thickens stellar populations by about the
  same proportion as the internal heating processes, for parts of the
  disc outside the bulge.}

In our simulations, the gas cooling function does not include metal
line or molecular cooling, limiting the temperature of the gas to
$>10^4\mathrm{~K}$.  As a result, the gas velocity dispersion is
largely set by the minimum temperature and consequently the gas scale
height increases as a function of radius. Due to this gas flaring, the
populations at each $\Delta R_g$ in Fig.~\ref{fig:zrms_jzfinal} are
actually born with different thicknesses, as shown in the middle
panel. The variation across the range of 5-6 Gyr old stars that end up
at the specified angular momentum is a factor $\sim2-3$. Hence, while
it is inevitable that individual populations change their vertical
distribution as they migrate (Fig.~\ref{fig:hz}), these changes are
partially offset by the flaring of stars at birth (middle panel of
Fig.~\ref{fig:zrms_jzfinal}). As a result, at a \emph{fixed} final
angular momentum (i.e. $R_g$) there is little variation of thickness
with $\Delta R_g$.

However, the molecular and atomic gas components in the MW show little
variation in thickness out to $\sim 10\mathrm{~kpc}$
\citep{Bronfman:1988,Wouterloot:1990,Narayan:2002}. We may therefore
expect that in the MW the flaring would not mask the thickening
resulting from radial migration. Evidence for similarly constant scale
heights as a function of radius has also recently been found with
resolved-star studies of edge-on nearby discs (R. de Jong \&
D. Streich, private communication).

We checked the results of this section against a higher-resolution
version of our simulation that uses four times as many particles
(simulation R4 from \citealt{Roskar:2012}) and found no significant
differences, so we consider the dynamical results to be numerically
robust. However, assessing the significance of the gas disc flaring
and the subsequent radial dependence of the scale height of young
stars is a much more difficult problem. It depends sensitively on the
details of the gas physics on scales that are an order of magnitude
smaller than what we (or any other similar state-of-the-art
simulations at present) are able to adequately resolve.

\section{Consequences for local abundance trends}
\label{sec:trends}

\begin{figure*}
\centering
\includegraphics[width=7in]{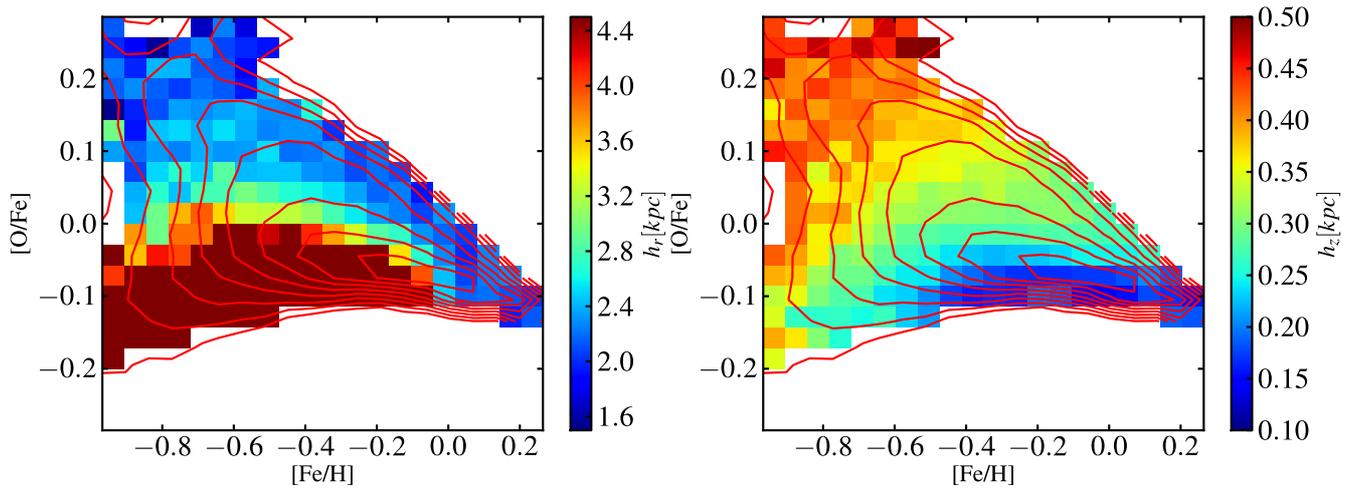}
\caption{The distribution of stars in the [O/Fe] vs. [Fe/H] plane,
  coloured by scale length $h_r$ on the left and scale height $h_z$ on
  the right. The logarithmically-spaced contours indicate mass density
  from 10-10$^4$ particles per bin. The stars lie between $4 < R
  \mathrm{~[kpc]} < 9$ and $|z| < 3 \mathrm{~[kpc]}$. }
\label{fig:ofe_feh}
\end{figure*}

The consequence of the thickening described above on the observed
properties of stellar populations is that their structural properties
vary considerably with metallicity and abundance. This has been
pointed out by \citet{Bovy:2012b}, who found that when they sliced the
SEGUE G-dwarf sample into ``mono-abundance'' bins, the structural
parameters varied smoothly from short and thick at the metal-poor
$\alpha$-rich end, to long and thin at the metal-rich $\alpha$-poor
end. We show a similar dissection of our model disc stars in
Fig.~\ref{fig:ofe_feh} and find a qualitatively similar result,
i.e. the scale height decreases smoothly from upper-left to
lower-right and the thinner populations tend to have longer radial
scale lengths. 

To obtain the structural parameters, we parametrize the stellar
distribution as 
\[
\rho(R,z) \propto f(R)f(z),
\]
where the radial and vertical density functions are
\[
f(R) = \mathrm{exp}\left(-\frac{R}{h_R}\right),
\]
and
\[
f(z) = \mathrm{sech}^2\left(-\frac{z}{2h_z}\right),
\]
respectively. We fit these functions to the simulated particle density
distribution for the maximum likelihood values of $h_r$ and $h_z$
using a procedure similar to the one described in
Sect.~\ref{sec:flaring}. We use stars within a range of $4 < R <
9$~kpc and $|z| < 3$~kpc to obtain the fits. By requiring at least 100
particles per each plotted cell, we obtain 1$\sigma$ uncertainties
that are everywhere less than 10\%. As in Sect.~\ref{sec:flaring}, we
experimented also with an exponential function for the vertical
distribution, and found that we can obtain smaller errors using the
sech$^2$ form. If we fit the vertical distributions of stellar
populations in our model with a single exponential, we obtain scale
heights that are somewhat higher than those shown in
Fig.~\ref{fig:ofe_feh}, similar to what we found in
Sect.~\ref{sec:flaring}.

The locus of ``saturated'' scale-lengths in Figure~\ref{fig:ofe_feh}
below $\mathrm{[O/Fe]}\sim -0.05$ are all young stars. Their
scale-lengths are long because they are forming out of the cold gas
which is also very extended. A similar saturation is seen in
\citet{Bovy:2012b}. In our model, the sub-solar part of this locus
originates in the outer disc, and \citet{Bovy:2012b} similarly find
that the mean radii for these stars are beyond the solar
neighborhood. These are young, 0-2 Gyr old stars that formed in the
outer disc and scattered into the solar neighborhood (see also
\citealt{Haywood:2008}). They did not migrate via the corotation
mechanism, because their tangential velocities exceed the circular
velocity by $\sim 20$~km/s. Therefore, they \emph{heated} and although
one would naively expect their negative $\Delta R$ to lead to a
thinner scale height, their scale heights are slightly higher than the
local young population by virtue of this heating. We can identify this
population in the lower right panel of Fig.~\ref{fig:hz} as the 2
Gyr-old stars that migrate inward by $\sim 1-2$~kpc and find that
their scale height $h_z \sim 0.3$~kpc.

An interesting discrepancy with the \citet{Bovy:2012b} distribution is
seen at the extreme metal-rich, $\alpha$-poor corner ([O/Fe] $< -0.1$
and [Fe/H] $> 0.1$). We find that those stars have a short
scale-length much like the older, more $\alpha$-rich population. In
\citet{Bovy:2012b} no such population exists. In our model, these
stars are dominated by intermediate age populations, mostly around
$\sim3$~Gyr and $5-6$~Gyr old.  The majority ($75\%$) of these stars
came from inside 3~kpc, i.e. from the same part of the disc as some of
the oldest stars in the solar neighborhood. These are the prototypical
``extreme'' migrators that were able to migrate far because they never
heated considerably. Hence, their overall velocity dispersion is $\sim
5-10$~km/s lower than other stars of similar age and because of the
lower heating they remain thinner than the average population of the
same age. Note that to increase particle numbers for more reliable
fitting, we extended the selection in Fig.~\ref{fig:ofe_feh} to
include stars to 4~kpc, this population still exists when we restrict
the sample to 7-9~kpc. We speculate that should such stars exist in
the MW, they may be absent from the SEGUE sample due to their thin
distribution and low numbers. However, should they be observed in
future surveys, they would be an indication of significant radial
mixing in the MW disc. An enticing connection to such a migrated
population may be the newly-discovered metal-rich, high-$\alpha$ stars
\citep{Adibekyan:2011, Adibekyan:2013, Gazzano:2013} whose chemical
properties suggest that they formed in the Galactic interior, but they
are kinematically akin to local thin-disc stars.

If the local abundance and metallicity distributions are affected by
stars migrating from the inner disc, one would expect to find
corresponding populations in the present-day bulge. The thickest parts
of the local thickened disc have low [Fe/H] and high $\alpha$
abundance, which may correspond qualitatively to a bulge population
`C' identified by \citet{Ness:2012}. The timing of the migration is
difficult to predict, though in general stars can migrate more easily
before they heat considerably. On the other hand the bulge region
seems to host stars of all abundances, metallicities, and ages
\citep{Bensby:2013}, limiting the constraints one could impose on the
models. In contrast to the local disc, however, the inner Galaxy does
show a strong age-metallicity relation. Therefore, if the thickened
population at the Sun's position is old and metal-poor, it could be
broadly consistent with being migrated from the inner
disc/bulge. Determining whether migration should be considered
critical for the evolution of the Milky Way will be determined in part
by upcoming chemical tagging surveys such as HERMES
\citep{Bland-Hawthorn:2010}.

\section{Conclusions}

We have demonstrated that an outward migrating population of stars in
a galactic disc \emph{always} thickens. At the same time, the vertical
velocity dispersion of such a population decreases, but we find that
despite this decrease the old stellar populations arriving from the
interior at the solar neighborhood match the kinematics of the MW
thick disc. Further, we find that radial migration and internal
heating thicken coeval stellar populations by comparable amounts. The
thickening due to radial migration alone is well-approximated by a
simple analytic treatment that assumes a conservation of vertical
action. Importantly, we find that while radial migration does cause an
increase in scale height with radius, the flaring that results is
minor. If radial migration can contribute to a thickened component of
a galactic disc then it has important consequences in the broader
context of disc galaxy formation since most other mechanisms proposed
to form a thick disc component involve the cosmological environment.

Our simulation recovers the qualitative structural stellar population
trends observed in the MW SEGUE data. In particular, we find a similar
dependence of scale height and scale length on oxygen abundance and
metallicity to those found in \citet{Bovy:2012b}. However, our model
produces a thick disc that is somewhat too thin compared to the
MW. This is to be expected, since the disc we considered in this work
is built up entirely from internal mechanisms in the absence of a
cosmological environment. Additional perturbations in a more realistic
setting would thicken the disc further.

Finally, we identify an interesting sub-population in the low-[O/Fe]
high-[Fe/H] corner of the abundance plane (Fig.~\ref{fig:ofe_feh}),
which appears to be absent from the \citet{Bovy:2012b} distribution,
perhaps due to the lack of low-latitude coverage in the SEGUE
data. These stars are the prototypical extreme-migrators having
migrated from $\sim3$~kpc in a few Gyr.

\section*{Acknowledgments}
R. R. thanks the Aspen Center for Physics and the NSF Grant \#1066293
for hospitality during the initial writing of this
manuscript. R. R. would also like to thank Greg Stinson, Jo Bovy,
Hans-Walter Rix, Glenn van de Ven, and Ralph Sch\"onrich for fruitful
discussions. R. R. is supported by the Marie Curie Career Integration
Grant and the Forschungskredit fellowship at the University of
Z\"urich.

\bibliographystyle{mn2e}
\bibliography{./disk_thickening.bib}

\end{document}